\documentclass[conference]{IEEEtran}
\IEEEoverridecommandlockouts

\usepackage{cite}
\usepackage{amsmath,amssymb,amsfonts}
\usepackage{amstext}
\usepackage{mathrsfs}
\usepackage{algorithm, algorithmicx}
\usepackage[noend]{algpseudocode}
\usepackage{graphicx}
\usepackage{float}
\usepackage{subfig}
\usepackage{textcomp}
\usepackage{xcolor}
\def\BibTeX{{\rm B\kern-.05em{\sc i\kern-.025em b}\kern-.08em
    T\kern-.1667em\lower.7ex\hbox{E}\kern-.125emX}}

\newcommand{\attackname}{F-LDDoS}

\newcommand{\Interval}{\ensuremath{\mathit{T}}}
\newcommand{\FeintingInterval}{\ensuremath{\mathit{F}}}
\newcommand{\AttackInterval}{\ensuremath{\mathit{A}}}
\newcommand{\AttackRate}{\ensuremath{\mathit{R}}}
\newcommand{\FeintingAttackRate}{\ensuremath{\mathit{{R}_{F}}}}
\newcommand{\SingleAttackRate}{\ensuremath{\mathit{{R}_{*}}}}
\newcommand{\Bandwidth}{\ensuremath{\mathit{C}}}
\newcommand{\Buffer}{\ensuremath{\mathit{B}}}
\newcommand{\BotSeries}{\ensuremath{\mathit{{S}_{bot}}}}
\newcommand{\AttackSeries}{\ensuremath{\mathit{{S}_{attack}}}}
\newcommand{\TotalSeries}{\ensuremath{\mathit{{S}_{all}}}}

\begin{document}

\title{Catch Me If You Can: A New Low-Rate DDoS Attack Strategy Disguised by Feint\\
}

\DeclareRobustCommand*{\IEEEauthorrefmark}[1]{%
    \raisebox{0pt}[0pt][0pt]{\textsuperscript{\footnotesize\ensuremath{#1}}}}

\author{
\IEEEauthorblockN{
Tianyang Cai\IEEEauthorrefmark{1}\IEEEauthorrefmark{3},
Yuqi Li\IEEEauthorrefmark{1}\IEEEauthorrefmark{3},
Tao Jia\IEEEauthorrefmark{1},
Leo Yu Zhang\IEEEauthorrefmark{2}, and
Zheng Yang\IEEEauthorrefmark{1}\IEEEauthorrefmark{*}}
\IEEEauthorblockA{\IEEEauthorrefmark{1}College of Computer and Information Science College of Software, Southwest University, Chongqing, China}
\IEEEauthorblockA{\IEEEauthorrefmark{2}School of Info Technology, Deakin University, Melbourne, Australia}
\IEEEauthorblockA{cai12345678@email.swu.edu.cn, lllyq13@email.swu.edu.cn, tjia@swu.edu.cn, leo.zhang@deakin.edu.au}
\IEEEauthorblockA{\IEEEauthorrefmark{*}Corresponding Author: Zheng Yang \quad Email: youngzheng@swu.edu.cn}
\IEEEauthorblockA{\IEEEauthorrefmark{3}Tianyang Cai and Yuqi Li are co-first authors.}}

\maketitle

\begin{abstract}
While collaborative systems provide convenience to our lives, they also face many security threats. One of them is the Low-rate Distributed Denial-of-Service (LDDoS) attack, which is a worthy concern.
Unlike volumetric DDoS attacks that continuously send large volumes of traffic, LDDoS attacks are more stealthy and difficult to be detected owing to their low-volume feature.
Due to its stealthiness and harmfulness, LDDoS has become one of the most destructive attacks in cloud computing.
Although a few LDDoS attack detection and defense methods have been proposed, we observe that sophisticated LDDoS attacks (being more stealthy) can bypass some of the existing LDDoS defense methods.
To verify our security observation, we proposed a new Feint-based LDDoS (\attackname) attack strategy.
In this strategy, we divide a Pulse Interval into a Feinting Interval and an Attack Interval. 
Unlike the previous LDDoS attacks, the bots also send traffic randomly in the Feinting Interval, thus disguise themselves as benign users during the {\attackname} attack.
In this way, although the victim detects that it is under an LDDoS attack, it is difficult to locate the attack sources and apply mitigation solutions.
Experimental results show that {\attackname} attack can degrade TCP bandwidth 6.7\%-14\% more than the baseline LDDoS attack.
Besides, {\attackname} also reduces the similarities between bot traffic and aggregated attack traffic, and increases the uncertainty of packet arrival.
These results mean that the proposed {\attackname} is more effective and more stealthy than normal LDDoS attacks.
Finally, we discuss the countermeasures of {\attackname} to draw the attention of defenders and improve the defense methods.

\end{abstract}

\begin{IEEEkeywords}
Collaborative System, Low-rate DDoS, Cloud Computing, Stealthy Attack Strategy
\end{IEEEkeywords}

\section{Introduction}

Collaboration system enables people to communicate and share documents through software and technology.
It includes many systems that are common in everyday life, such as email systems, web systems, multimedia systems, and cloud computing systems.
However, these public service systems are often targeted by attackers because of their importance.
For example, there are many security issues in cloud computing, such as insecure APIs, hijacking of accounts, malicious insiders, and denial of service.
One of these, known as Low-rate denial-of-service (LDoS) attack~\cite{kuzmanovic2003low}, is regarded as one of the most harmful attack techniques in cloud computing~\cite{agrawal2019defense, zhijun2020low}.

\begin{figure}
    \centering
    \includegraphics[width=0.8\columnwidth]{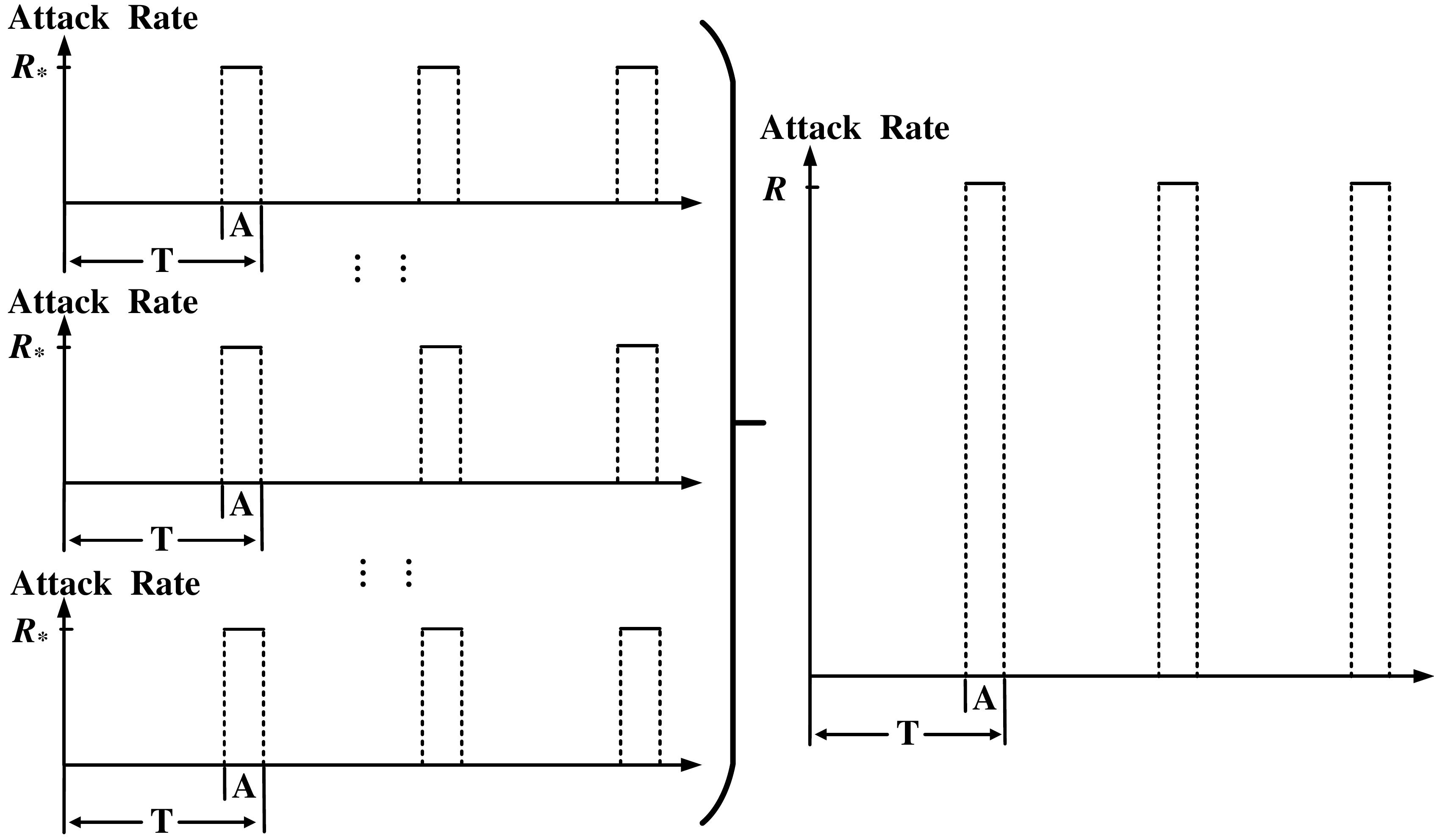}
    \caption{Low-rate Shrew DDoS attack.}
    \label{fig:LDDoS}
    \vspace{-3mm}
\end{figure}

An LDoS attack aims to dramatically reduce the victim's transmission performance by lowering the target TCP bandwidth by taking advantage of the TCP retransmission timeout (RTO)~\cite{padhye1998modeling} mechanism's weakness.
Specifically, an LDoS attacker can make the bottleneck link periodically congested by injecting high-density impulse traffic to trigger the victim's TCP congestion control mechanism, and therefore reduce TCP bandwidth. 
Compared with volumetric DDoS attacks, LDoS attacks are less costly but more difficult to be detected.
As the attacks continue to evolve, single-source LDoS attacks have evolved into low-rate distributed denial-of-service (LDDoS) attacks.
The schematic diagram of the LDDoS attack is shown in Figure~\ref{fig:LDDoS}. 
Similar to other DDoS attacks, LDDoS is launched by a group of bots. 
These bots send small pulse traffic synchronously, which converge to form huge periodic pulses, resulting in an LDDoS attack.

Recently, various defense methods for LDoS and LDDoS have been proposed (e.g.,~\cite{tang2021performance, csimcsek2018fast, tang2022novel, tang2020fr, tang2021mf, tang2022new, wang2021ldos, shi2022approach,ilango2022feedforward, wu2019sequence, wu2015low}).  
According to the features they use, they can be classified into methods based on network statistical features~\cite{csimcsek2018fast, tang2021mf, tang2022new, wang2021ldos, shi2022approach,ilango2022feedforward}, and methods based on traffic rate sequence~\cite{tang2022novel, tang2021performance, tang2020fr, wu2019sequence, wu2015low}.
All of them show good detection performance for specific LDoS and LDDoS attacks. 
However, despite the many detection methods already available, mitigating LDDoS attacks is still a hard nut to crack. 
This is because, although both exhibit the characteristic of pulse traffic, LDDoS is more difficult to defend against than LDoS attacks due to its distributive nature. 

As a representative defense based on traffic rate sequence, Tang \textit{et al.}~\cite{tang2021performance} proposed a method to locate the LDoS attacker based on the similarity of traffic rate sequences.
This idea can be generalized to mitigate LDDoS attacks by assuming that the attack traffic of a single bot and the aggregated traffic are still highly similar.
But it raises \textbf{Question 1:} 
whether LDDoS attacks are harder to defend in a (realistic) setting where the single attack flows are not similar to the aggregated attack flow? 
As a typical example of network statistical feature-based defense, authors in~\cite{csimcsek2018fast} proposed an LDDoS filtering approach. 
This method takes advantage of the periodic nature of the bot's traffic and calculates the standard deviation of the packet arrival time intervals.
Since the LDDoS attack has a pulse interval, the standard deviation of the bot's flow is significantly lower than that of benign user.
However, it raises \textbf{Question 2:} will this method fail if the arrival time intervals of the bot's packets are random rather than exhibiting periodicity? 

\begin{figure}
    \centering
    \includegraphics[width=0.8\columnwidth]{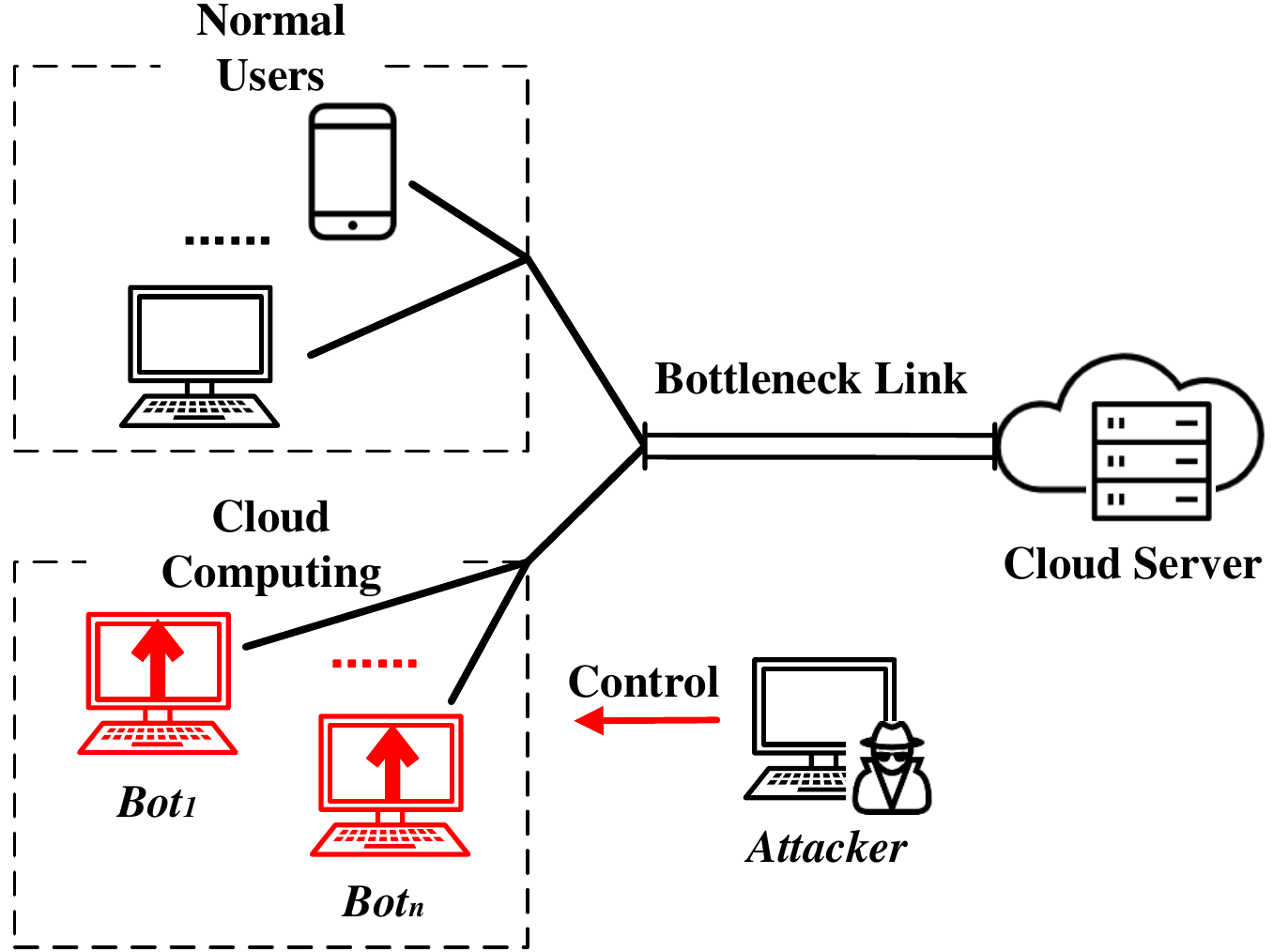}
    \caption{LDDoS attack scenario targeting cloud computing.}
    \label{fig:scenario}
\end{figure}

\textbf{Our work.} While detection and defense methods are important research problems, assessing the feasibility and limitations of LDDoS attack strategies under realistic assumptions is also an essential challenge. 
Therefore, to address the above questions, we propose a novel Feint-based LDDoS ({\attackname}) attack strategy that uses feint traffic to make the single attack flows more stealthy, thus making it more difficult to defend against.
In this strategy, we divide the Pulse Interval into Feinting Interval and Attack Interval.
Different from normal LDDoS attacks, the bots also send traffic randomly in the Feinting Interval and disguise themselves as benign users.
Therefore, when multiple seemingly harmless flows converge, they can form powerful periodic pulse traffic, thus causing LDDoS attacks.
With this approach, we not only address (\textbf{Question 1}) by reducing the similarity between the traffic of a single bot and the total attack traffic but also address (\textbf{Question 2}) by making the arrival time intervals of the bot's packets random.
This means that it is difficult for the victim to filter out {\attackname} attack traffic, even though it detects that it is under LDDoS attack.
We evaluate the proposed attack strategy's effectiveness and stealthiness by comparing it to the normal LDDoS attack as a baseline.
Our experimental results show that {\attackname} degrades TCP bandwidth more (6.7\%-14\%), even though its attack cost is lower than the normal LDDoS attack.
Moreover, we increase the Dynamic Time Warping distance (i.e., decreased in similarity) between bot traffic and aggregated attack traffic by about 50\% to 70\%, and we greatly increase the standard deviation of the packet arrival time intervals.
These results show that the {\attackname} attack is more effective and more stealthy than the baseline LDDoS attack.

\textbf{Paper Outline.}
We model attack scenarios for bottleneck links in cloud computing in Section~\ref{scenarios}.
Then, we present the detailed design of {\attackname} attack strategy in Section~\ref{design}.
In Section~\ref{evaluation}, we evaluate the proposed strategy in effectiveness and stealthiness.
After that, we discuss the countermeasure of {\attackname} in Section~\ref{discussion}.
And we discuss some related works in Section~\ref{related_work}.
Finally, we conclude this paper in Section~\ref{conclusion}.

\section{Modeling Attack Scenarios for bottleneck links in Cloud Computing}\label{scenarios}

In order to achieve a successful LDoS attack, the following requirements need to be satisfied.
\begin{itemize}
    \item 
    To produce bottleneck link congestion, the attack rate {\AttackRate} must be higher than the link bandwidth {\Bandwidth} or equal to it.
    \item 
    Attack interval {\AttackInterval} needs to be either longer than RTT or long enough to fill link buffer {\Buffer}. However, {\AttackInterval} must not be excessively lengthy to prevent being recognized as a high-rate DDoS attack.
    \item The pulse interval {\Interval} must be greater than or equal to the initial retransmission timeout (RTO) of the target TCP sender.
\end{itemize}

As the attacks continue to evolve, single-source LDoS attacks have evolved into distributed LDDoS attacks.
The LDDoS attack scenario targeting cloud computing is shown in Figure~\ref{fig:scenario}.
For LDDoS attacks, the following conditions should also be considered because the delay between bots affects the effectiveness of the attack.
\begin{itemize}
    \item The communication delay between the attacker and bots. 
    If these delays are not small enough, the time between these bots launching the attack will be out of sync.

    \item The asynchrony in the convergence of different attack flows. If the attack flows take different paths, they may reach the bottleneck link at different moments.
\end{itemize}

These delays can lead to shorter attack intervals and lower peaks after convergence, thus reducing the effectiveness of LDDoS attacks.

\section{Design of {\attackname} Attack Strategy}\label{design}
The proposed attack strategy, which complies with the demands in Section~\ref{scenarios}, is described in depth in this section.

\begin{figure}
    \centering
    \includegraphics[width=0.8\columnwidth]{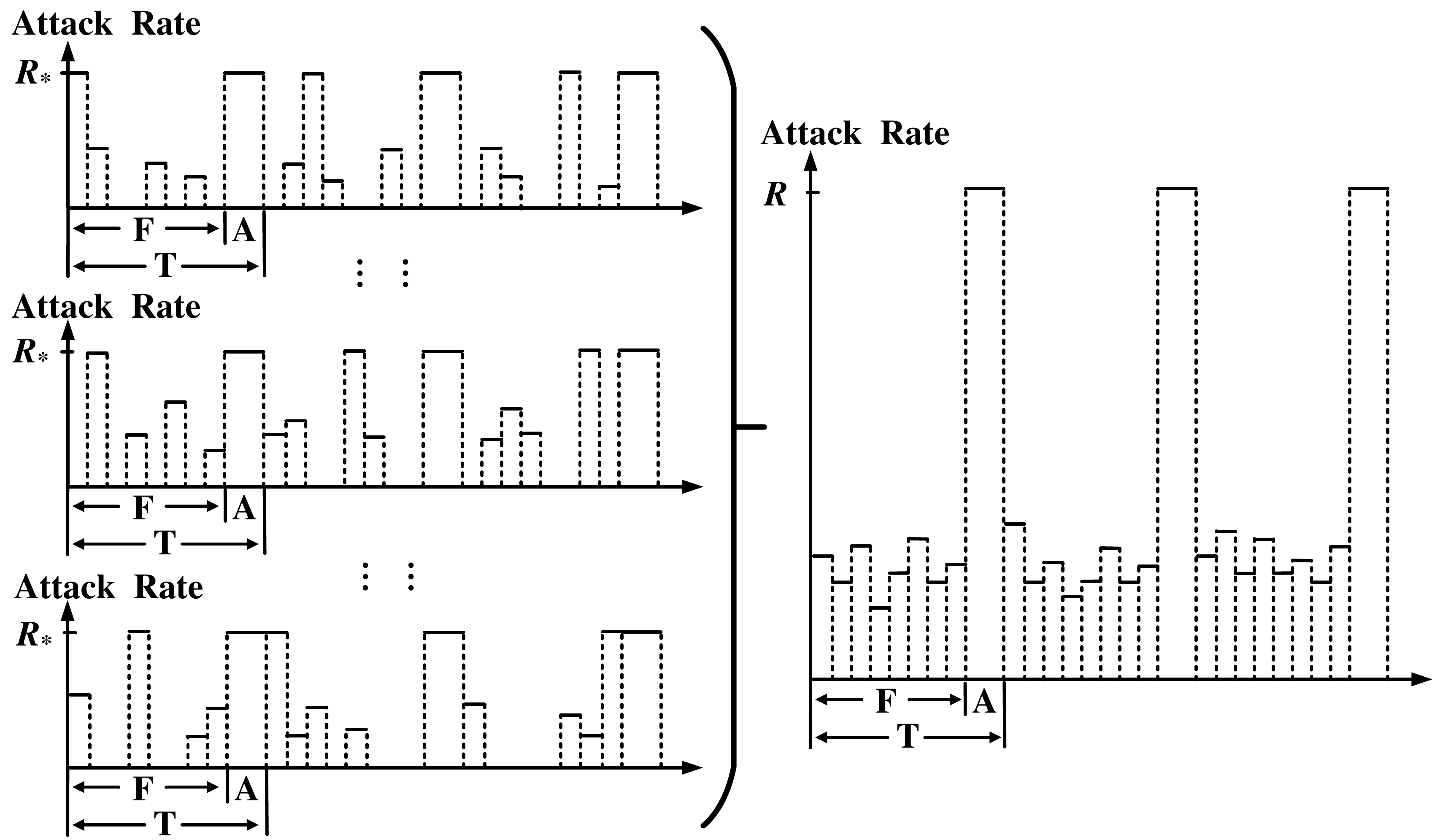}
    \caption{Schematic diagram of proposed {\attackname} attack.}
    \label{fig:attack_idea}
\end{figure}

Compared to the classical Shrew LDDoS attack~\cite{kuzmanovic2003low}, the main goal of our proposed attack is to be more \textit{stealthy}.
That is, our attack should be more difficult to detect and mitigate.
But in a normal LDDoS attack, the attack waves of single bot traffic and aggregated attack traffic are highly similar, both act as periodic square waves (as shown in Figure~\ref{fig:LDDoS}).
To address this limitation, our main idea for designing {\attackname} is that the traffic of each bot is seemingly harmless, but when aggregated, it creates huge traffic pulses.
As shown in Figure~\ref{fig:attack_idea}, the {\attackname} attack is generated from a group of bots, and each of their attack waves behaves like a normal user.
When these flows converge together, they form a periodic pulse wave, resulting in an LDDoS attack.
Thus, although the victim knows that it is under an LDDoS attack, it is difficult to identify which flows contribute to the attack and thus cannot filter them accurately.

\subsection{Attack Strategy}
\label{Attack_Strategy}
For ease of understanding, we define the Pulse Interval as {\Interval}.
And a {\Interval} includes a Feinting Interval ({\FeintingInterval}) and an Attack Interval ({\AttackInterval}).
The relationship between the three of them is: 
\begin{equation}
\Interval = \FeintingInterval + \AttackInterval.
\end{equation}

In the Feinting Interval, as its name implies, each bot sends small random attack flows, thus disguising itself as a normal user as much as possible.
The range of attack rates for a single bot is $[0, \SingleAttackRate]$, and the bot sends attack traffic at random moments in this interval. 
That is, at each moment of each Feinting Interval, there is a 50\% probability of sending attack traffic and the average value of the attack rate is $\SingleAttackRate \times 50\%$.
Thus, assuming that there are $m$ bots, the aggregated feinting attack rate {\FeintingAttackRate} is calculated as:
\begin{equation}
\FeintingAttackRate = \sum_{i=1}^{m} \SingleAttackRate_{i} \approx m \times 50\% \times \SingleAttackRate \times 50\% = 0.25m\SingleAttackRate.
\end{equation}

However, in the Attack Interval, all bots send attack traffic simultaneously, thus causing congestion on the target bottleneck link. 
The traffic rate of a single bot is {\SingleAttackRate}.
Therefore, the aggregated attack rate {\AttackRate} is calculated as:
\begin{equation}
\AttackRate = m \times\SingleAttackRate. 
\end{equation}

\subsection{Attacker-Bots Synchronization}

The communication between the attacker and different bots may have different time delays.
If the attacker sends a command to all bots to launch an attack immediately, the time when each bot starts sending attack traffic may not be synchronized due to the different time delays.
As a result, the Attack Intervals of each bot may not overlap completely, making the actual Attack Interval shorter and the peak rate lower, which severely degrades the attack effectiveness.
To avoid this problem, we consider these two aspects: asynchrony of sending attack traffic and attack traffic aggregation.

To reduce the asynchrony of sending attack traffic, we specify the time to launch the attack in the command.
For example, the attacker sends a command to all bots at 18:59:30 to launch the attack at 19:00:00.
In this way, even though the bots may receive the command at different moments, they will all launch the attack simultaneously.

\begin{algorithm}[t]
\caption{{\attackname} attack implementation.}
\label{alg_implementation}
\begin{algorithmic}[1]
    \Require {\Interval} \Comment{Pulse Interval (s).}
    \Require {\AttackInterval} \Comment{Attack Interval (s), $0<\AttackInterval \le \Interval$.}
    \Require {\AttackRate} \Comment{Rate of aggregated attack traffic (Mbps).}
    \Require $m$ \Comment{Bot amount.}
    \Require $D$ \Comment{Attack Start Time.}
    \Require $IP$ \Comment{Victim's IP address.}
    \Require $Port$ \Comment{Target port.}
    \State {\FeintingInterval} = {\Interval-\AttackInterval} \Comment{Feinting Interval (s).}
    \State {\SingleAttackRate} = {$\AttackRate / m$} \Comment{Attack rate for a single bot (Mbps).}
    \State \textbf{\# Step 1:} Check the time of the attack.
    \While{True} \Comment{Check whether to start the attack.}
        \State CurrentTime = datetime.now() 
        \If{CurrentTime $\ge D$}
            \State \textbf{break}
        \EndIf
    \EndWhile
    \While{True} \Comment{Start attack.}
    \State \textbf{\# Step 2:} Generate attack series.
        \State AttackSeries = []
        \For{i in range($\FeintingInterval/0.1$)} \Comment{For each $0.1$s in {\FeintingInterval}.}
            \If{random()$> 0.5$} 
                \State AttackSeries.append({\SingleAttackRate} $*$ random())
            \Else
                \State AttackSeries.append($0$)
            \EndIf
        \EndFor
        \For{i in range($\AttackInterval/0.1$)} \Comment{For each $0.1$s in {\AttackInterval}.}
            \State AttackSeries.append({\SingleAttackRate})
        \EndFor
        \State \textbf{\# Step 3:} Send attack traffic.
        \State s = socket(AF\_INET, SOCK\_DGRAM)
        \State LastTime = datetime.now()
        \For{i in range($\Interval/0.1$)} 
            \While{True} \Comment{Send attack traffic in each $0.1$s.}
                \State CurrentTime = datetime.now()
                \If{CurrentTime - LastTime $<100$ms}
                    \State \textbf{continue}
                \Else
                    \State LastTime = CurrentTime
                    \State Amount = AttackSeries[i]$*10^6/m/8/80$
                    \State Payload = `0'$*80$
                    \For{j in range(Amount)}
                        \State s.sendto(Payload.encode(), ($IP$,~$Port$))
                    \EndFor
                    \State \textbf{break}
                \EndIf
            \EndWhile
        \EndFor
    \EndWhile  
\end{algorithmic}
\end{algorithm}

To reduce the asynchrony of attack traffic aggregation, we can only make the bots deploy in close networks because the path of the attack traffic is not known.
In particular, we allow all bots to execute on a virtual machine created using the cloud computing service in an area close to the target.
By allowing the attack traffic to follow the same attack path as much as feasible, this minimizes the asynchrony associated with the use of alternative routes.

\begin{figure}
    \centering    \includegraphics[width=0.8\columnwidth]{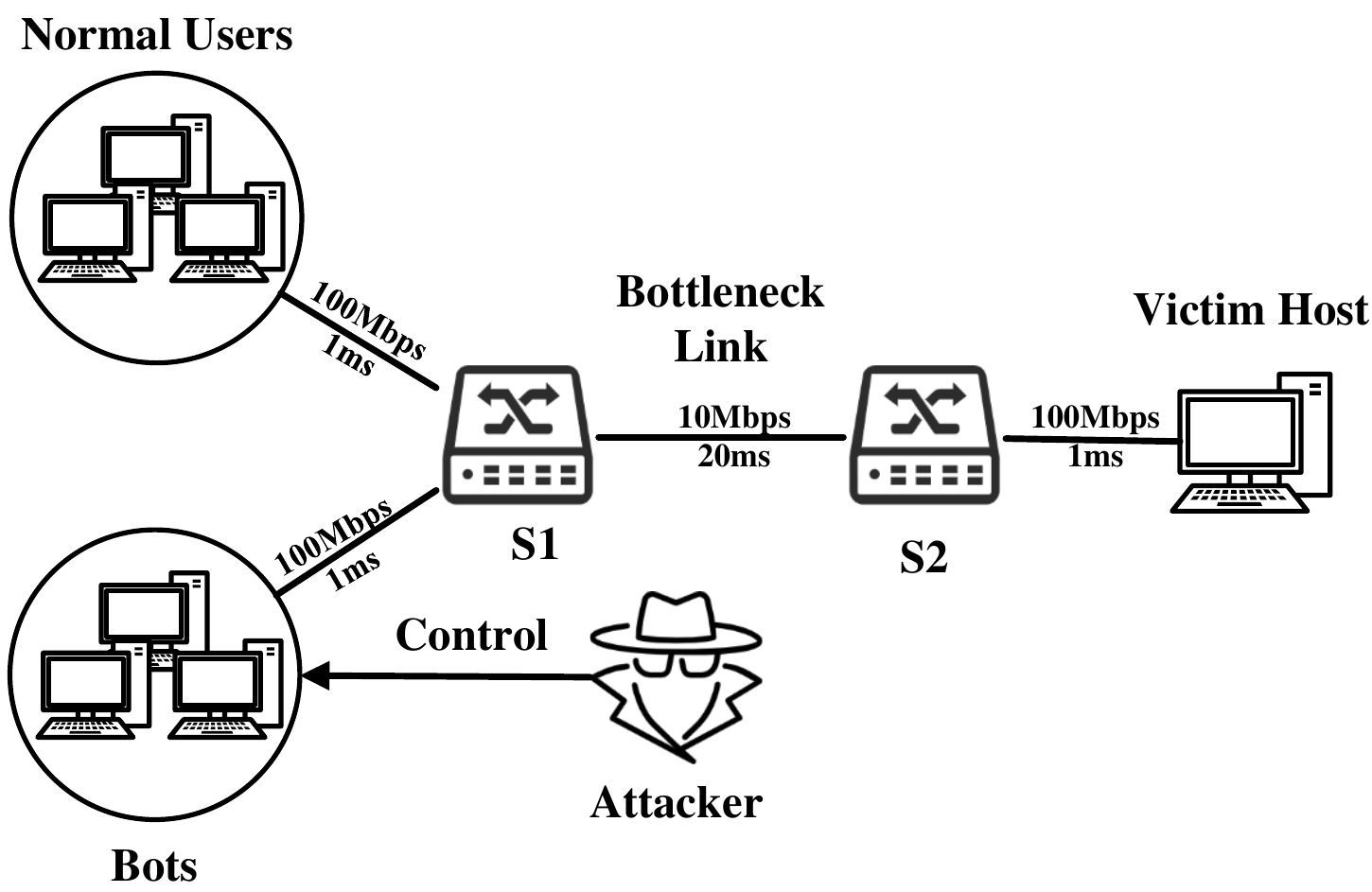}
    \caption{Experiment topology.}
    \label{fig:topo}
\end{figure}

\subsection{{\attackname} Attack Implementation}
\label{implementation}
The detailed implementation of the {\attackname} attack is shown in Algorithm~\ref{alg_implementation}.
This implementation can be divided into three steps that are described as follows.
\begin{itemize}
    \item Step 1: Once the bot receives a command, it then starts cycling to check whether the attack time is currently reached. 
    If the current time is greater than or equal to the attack time, the bot immediately enters the next step.
    
    \item Step 2: In this step, the bot generates an attack series for a {\Interval} according to the attack strategy proposed in Section~\ref{Attack_Strategy}. 
    And the attack series indicates the attack rate of the bot every 0.1s in {\Interval}.
    
    \item Step 3: After the attack series is generated, the bot starts to construct and send attack packets in each 0.1s.
    The size of each UDP packet is 80 bytes, and the amount of packets is calculated through the attack series.
    When the attack traffic for the current interval {\Interval} is sent, the bot returns to Step 2 and thus sends the attack traffic cyclically.
\end{itemize}

\section{Evaluation of {\attackname} Attack}\label{evaluation}
In this section, we evaluate {\attackname} attack strategy from the aspects of effectiveness and stealthiness compared with baseline LDDoS attack.

\subsection{Experiment Setup}
The experiment environment is set up on a computer with 16GB of RAM, Intel Core i5-7500 CPU, and Ubuntu 20.04.1. 
We utilize Mininet~\cite{Mininet} as the network simulator and the topology is shown in Figure~\ref{fig:topo}. 
Between switches S1 and S2, there is a bottleneck link with a 10Mbps bandwidth and a 20ms delay.
Additionally, other links have a 100Mbps bandwidth and a 1ms latency.
S1 connects a group of normal users who have established TCP connections with the victim host that connects to S2.
And S1 also connects a group of bots that send UDP-based attack traffic to the victim host.
Reno is the setting for the TCP congestion control mechanism.
The other parameters of TCP protocol are set to the default value of the Python socket.

In the experiment, the traffic in the network topology is in two parts: 
\begin{itemize}
    \item The TCP traffic sent by normal users to the victim. 
    We achieve this traffic by using TCP to transfer a large file. 
    Since there is no speed limit when sending the file, we assume that the TCP flow rate in the bottleneck link is representative of the maximum bandwidth of TCP.

    \item The UDP attack traffic sent by bots to the victim.
    We achieve this traffic through the attack implementation proposed in Section~\ref{implementation}.
    Since the principle of the attack is to make the bottleneck link periodically congested, we use UDP traffic as the attack traffic in order to facilitate the measurement of TCP bandwidth degradation.
\end{itemize}

\subsection{Attack Parameter Setting}
The minimum RTO, which is the initial RTO of the TCP sender, must not be less than the pulse interval {\Interval} or equal to it.
In the following tests, {\Interval} is set to $1$ second since RFC6298 advises that the minimum RTO be set to $1$ second~\cite{paxson2011computing}.
The bot amount $m$ is set to $10$, which means we use $10$ bots to launch {\attackname} attack.
The Attack Interval {\AttackInterval}, Feinting Interval {\FeintingInterval}, and attack rate {\AttackRate} are set to different values in the following experiments.

\begin{figure*}[!t]
    \centering
    \subfloat[{\attackname}, $\AttackRate=10$Mbps, $\AttackInterval=0.1$s, $\FeintingInterval=0.9$s.]{\includegraphics[width=0.4\textwidth]{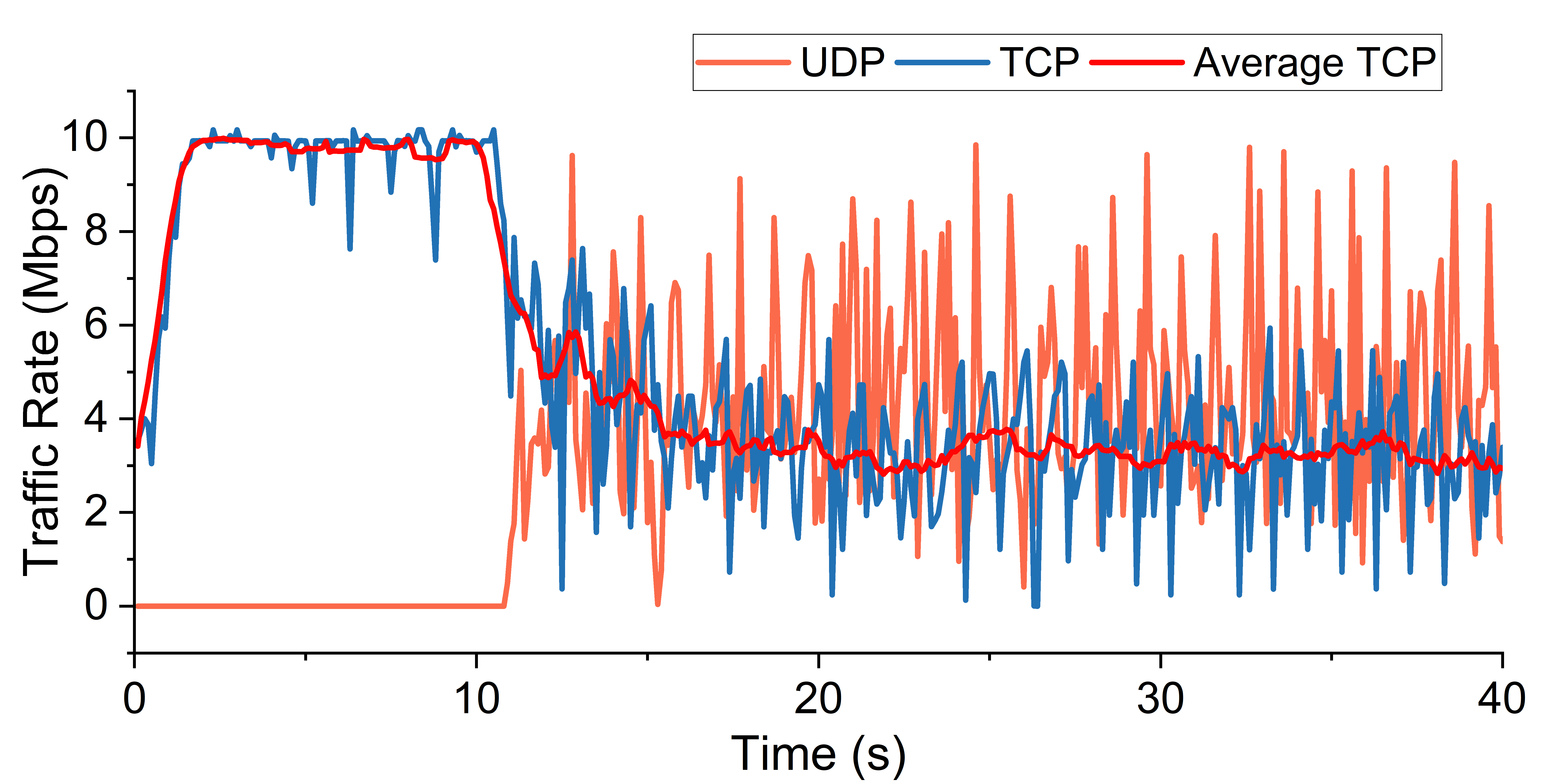}%
    \label{fig_s_10_0.1}}
    \subfloat[{\attackname}, $\AttackRate=10$Mbps, $\AttackInterval=0.2$s, $\FeintingInterval=0.8$s.]{\includegraphics[width=0.4\textwidth]{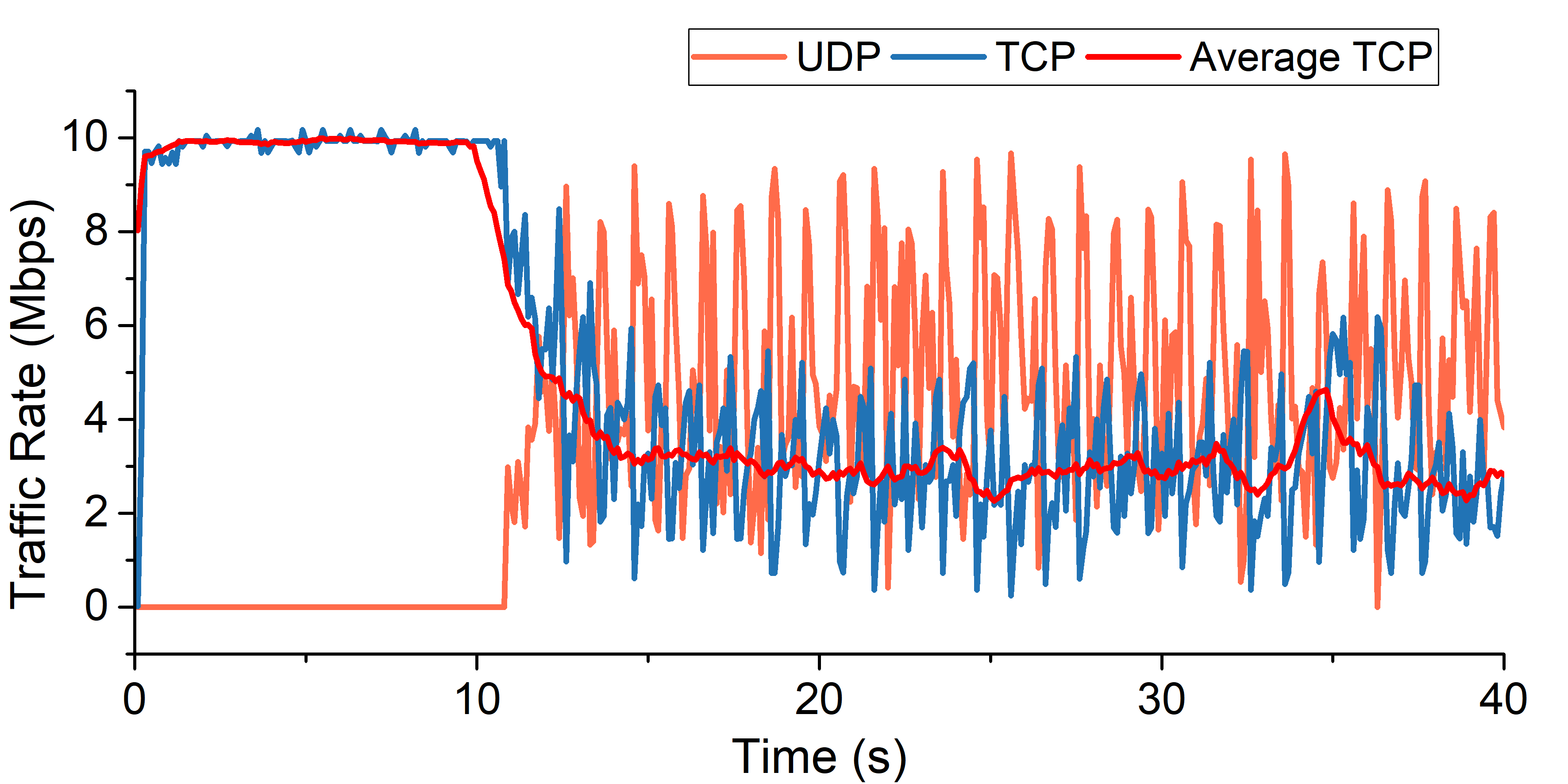}%
    \label{fig_s_10_0.2}}
    
    \subfloat[LDDoS, $\AttackRate=10$Mbps, $\AttackInterval=0.2$s.]{\includegraphics[width=0.4\textwidth]{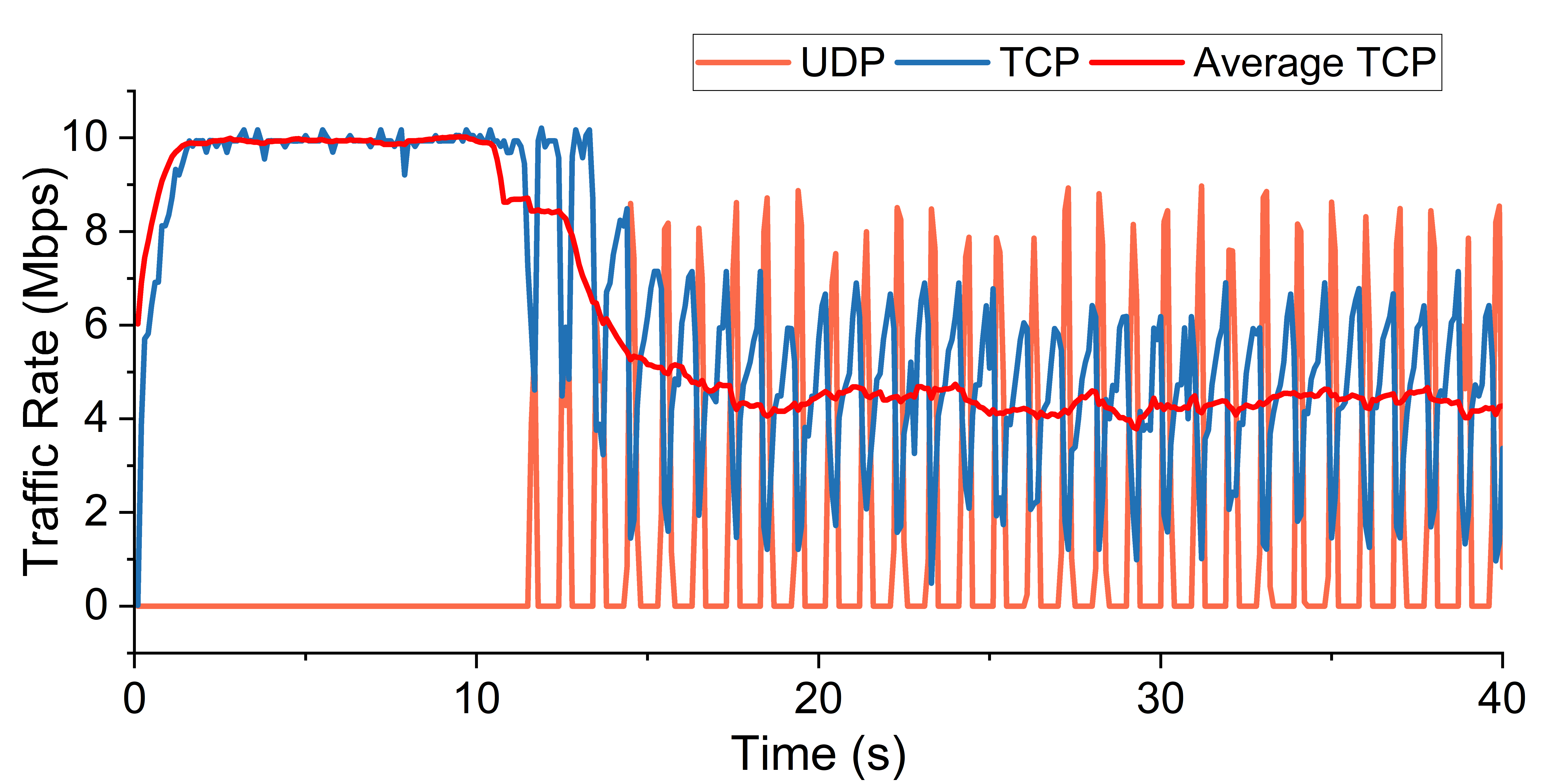}%
    \label{fig_n_10_0.2}}
    \subfloat[LDDoS, $\AttackRate=20$Mbps, $\AttackInterval=0.2$s.]{\includegraphics[width=0.4\textwidth]{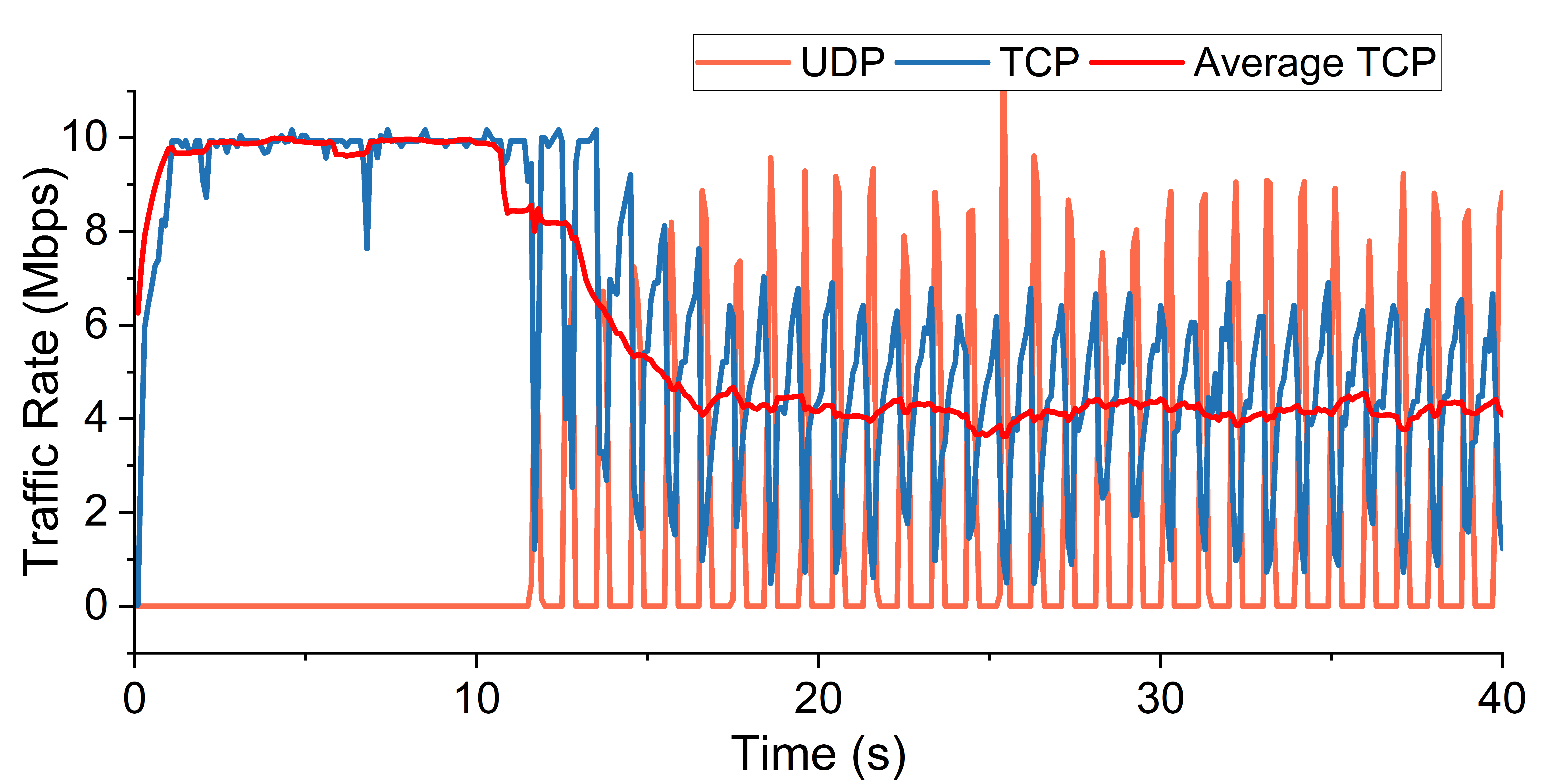}%
    \label{fig_n_20_0.2}}
    \caption{Attack effectiveness of {\attackname} and LDDoS.}
    \label{fig:waves}
\vspace{-3mm}
\end{figure*}

\begin{table}[]
\caption{Effectiveness and stealthiness evaluation results of {\attackname} and LDDoS attacks.}
\label{table:results}
\setlength{\tabcolsep}{2pt}
\resizebox{\columnwidth}{!}{%
\begin{tabular}{ccccc}
\hline
Attack Type & \multicolumn{2}{c}{\attackname} & \multicolumn{2}{c}{LDDoS (baseline)} \\ \hline
Attack Interval {\AttackInterval} & 0.1 s & 0.2 s & 0.2 s & 0.2 s \\
Feinting Interval {\FeintingInterval} & 0.9 s & 0.8 s & None  & None \\
Attack Rate {\AttackRate} & 10 Mbps & 10 Mbps & 10 Mbps & 20 Mbps \\
Attack Cost & 3.25 Mbps & 4 Mbps & 2 Mbps & 4 Mbps \\
Attack Peak Rate & 9.85 Mbps & 9.67 Mbps & 9.21 Mbps & 12.92 Mbps \\
TCP Bandwidth after Attack & 3.44 Mbps & 2.92 Mbps & 4.37 Mbps & 4.14 Mbps \\
TCP Bandwidth Decline Ratio & 65.98\% & 71.28\% & 57.20\% & 59.29\% \\
DTW Distance of {\BotSeries} and {\AttackSeries} & 95.46 & 93.36 & 62.30 & 55.81 \\
DTW Distance of {\BotSeries} and {\TotalSeries} & 184.71 & 165.38 & 134.74 & 128.77 \\ 
$\sigma$ of packet arrival time interval & 0.1113 & 0.1083 & 7.8*$10^{-16}$ & 8.8*$10^{-16}$ \\ \hline
\end{tabular}%
}
\end{table}

\subsection{Effectiveness Evaluation}
To validate the effectiveness of {\attackname} attack, we launch the attack and measure the TCP and UDP traffic rates.
Specifically, we use Wireshark~\cite{wireshark} to capture all packets at S2, then we calculate the TCP and UDP traffic rate per $0.1$ second.

We launch the {\attackname} with $\AttackRate=10$Mbps, $\AttackInterval=0.1$s and $0.2$s.
For comparison, we launch the normal LDDoS with $\AttackRate=10$Mbps and $20$Mbps, $\AttackInterval=0.2$s.
The traffic rates during different attacks are shown in Figure~\ref{fig:waves}.
The red lines in Figure~\ref{fig:waves} represent the average TCP bandwidth in $1$ second.
As we can see from the figures, both proposed {\attackname} and LDDoS can effectively degrade the TCP bandwidth through periodic pulse traffic.
And the TCP bandwidth under {\attackname} attack is lower than LDDoS.

To further quantitatively compare the {\attackname} and LDDoS attacks, we calculate their attack costs, attack peak rates, TCP bandwidth after attack, and TCP bandwidth decline ratios respectively in Table~\ref{table:results}.
The attack cost is given by $\AttackRate\times\AttackInterval+\FeintingAttackRate\times\FeintingInterval=\AttackRate(\AttackInterval+0.25\FeintingInterval)$.
As we can see from the table, {\attackname} achieves greater TCP decline ratios than LDDoS, in both larger and the same attack cost.
Even if the {\AttackInterval} is as short as $0.1$s, our {\attackname} still achieves 65.98\% TCP bandwidth decline, which is still larger than LDDoS.
With the introduction of feint traffic, there is an inevitable increase in the attack cost.
However, since the attack is distributed, we believe that the cost of the attack spread over each bot is acceptable.

\subsection{Stealthiness Evaluation}
Stealthiness is the main goal of the proposed {\attackname} attack. 
Our proposed attack, although detectable, would make it difficult for the victim to distinguish which flows are the attack flow and thus unable to filter the attack traffic. 
We define the attack rate series of a single bot as {\BotSeries}, the aggregated attack rate series as {\AttackSeries}, and the total traffic rate series as {\TotalSeries}.
Therefore, the goal of our attack strategy is to reduce the similarity between {\BotSeries} and {\AttackSeries}.
In addition, reducing the similarity between {\BotSeries} and {\TotalSeries} can also increase the difficulty of attack traceability and mitigation.
Besides, since the attack packet arrival time interval of bots could also be utilized to filter attack traffic, we need to verify whether the proposed strategy decreases the stability of the attack packet arrival time interval of bots.

To measure the similarity between series, we use Dynamic Time Warping (DTW), which is also used to locate attackers in ~\cite{tang2021performance}.
DTW can compare series that are unequal in length.
The larger the DTW distance between two series, the less similar they are. 
Conversely, the smaller the DTW distance, the higher their similarity.
We measure the DTW distance between {\BotSeries} and {\AttackSeries}, and {\BotSeries} and {\TotalSeries} in Table~\ref{table:results}.
We can find that the {\BotSeries}-{\AttackSeries} DTW distance of {\attackname} is larger than LDDoS.
For instance, both with $\AttackInterval=0.2$s and $\AttackRate=10$Mbps, the {\BotSeries}-{\AttackSeries} DTW distance of {\attackname} increased 50\% compared to LDDoS.
Moreover, with equal attack costs (both in $4$Mbps), this distance of {\attackname} increased 67.7\% compared to LDDoS.
Besides, the {\BotSeries}-{\TotalSeries} DTW distance of {\attackname} also increases from $22.7\%$ to $43.4\%$ compared to LDDoS.

To measure the stability of packet arrival time intervals, we calculate the standard deviation $\sigma$ of the packet arrival time interval, which is also used to determine attack flows in~\cite{csimcsek2018fast}.
The results are shown in Table~\ref{table:results}.
We can find that $\sigma$ of packet arrival time interval in {\attackname} is greatly larger than LDDoS.
According to~\cite{csimcsek2018fast}, benign users' $\sigma$ is larger than $1\times10^{-4}$, meaning that the bots of {\attackname} have successfully disguised themselves as benign users. 
These results show that our proposed {\attackname} attack is more stealthy and harder to defend than the regular LDDOS attacks.

\section{Discussion of {\attackname} Countermeasure}
\label{discussion}
We consider two directions to defend the proposed attack.
One of them is to identify and adjust the bottleneck links in the network. 
An attacker needs to probe the bottleneck link before launching an LDDoS attack.
If the network can dynamically identify the bottleneck link and adjust the routing or forwarding rule to avoid congestion, this will cost the attacker more time and resources.
Another one is to customize a defense method against {\attackname}.
Since {\attackname} can be detected, the defender can get its Pulse Interval {\Interval} and Attack Interval {\AttackInterval}.
Therefore, the defender can check which flows always send {\AttackInterval} seconds of traffic with a {\Interval} period, and identify these flows as attack flows.
These methods can be implemented based on SDN, which has a global view of topology and can access flow-level statistics.

\section{Related Work}
\label{related_work}
In 2020, Park \textit{et al.}~\cite{park2020assessing} assessed the feasibility of Very Short Intermittent DDoS attacks (VSI-DDoS).
This attack assumes that the attack traffic arrives at the victim server within a very short time interval.
The results demonstrated that VSI-DDoS attacks are asynchronization vulnerable.
When the synchronization delay is $90$ms, the effect of the attack is already reduced by $85.7\%$.
Since the LDDoS attack also causes a decrease in attack effectiveness in the case of asynchrony, it is important to evaluate the effectiveness and stealthiness of {\attackname} attack with different delays.

In 2021, Takahashi \textit{et al.}~\cite{takahashi2021low} proposed an LDDoS approach that targets residential networks with unknown bottleneck link characteristics.
This strategy can automatically adjust the attack rate depending on the number of compromised devices in the residential network.
Therefore, the attacker may still successfully execute an LDDoS attack even if he is unaware of the bandwidth and latency of the bottleneck link.

In 2021, Tang \textit{et al.}~\cite{tang2021performance} proposed an SDN-enabled LDoS detection and mitigation approach called P\&F. 
This approach takes advantage of the SDN architecture and enables easier collection of flow-level statistics.
It detects LDoS attacks based on performance and features. 
Moreover, it mitigates LDoS attacks based on similarities between aggregated flow and single flow to locate the attacker. 
And it inspired us to design {\attackname} to bypass this mitigation method.

In 2021, Yue \textit{et al.}~\cite{yue2021high} explored high-potency patterns of the LDoS attack and developed two attack models to maximize attack potency.
This work targets on CUBIC TCP congestion control algorithm and RED queue management scenario, and aims to improve the attacking potency.
In contrast, our work mainly focuses on increasing the stealthiness of LDDoS attacks.
However, we believe that these two works may complement each other to construct more destructive attack methods.

\section{Conclusion}
\label{conclusion}
In this paper, we propose a new Feint-based low-rate DDoS attack strategy (named {\attackname}) to increase the difficulty of detection and mitigation.
The key idea of {\attackname} is to make the traffic sent by bots seem harmless, but when they are aggregated, they form a powerful LDDoS attack.
Specifically, we divide the Pulse Interval {\Interval} into Feinting Interval {\FeintingInterval} and Attack Interval {\AttackInterval}.
And we let bots send feint traffic randomly in the F interval, and disguise themselves as normal users.
Experimental results show that {\attackname} not only has better performance on TCP degradation but is also more stealthy than normal LDDoS attacks.

In the future, we will validate our attack strategy using more complex experimental scenarios, including complex network topologies, complex network parameters, and simulated background traffic.
In addition, we will use the state-of-the-art LDDoS defense systems to validate the effectiveness and stealthiness of our strategy.

\section{Acknowledgment}


This work is supported by the Natural Science Foundation of Chongqing (Grant No. CSTB2022NSCQ-MSX0437) and University Innovation Research Group of Chongqing (Grant No. CXQT21005).

\bibliographystyle{IEEEtran}

\bibliography{main}

\end{document}